\input harvmac
\overfullrule=0pt
%\draft
\Title{\vbox{%\baselineskip12pt
\hbox{USC-96/18}
\hbox{hep-th/9608047}
}}
{\vbox{\centerline{An $N=2$ Superconformal Fixed Point}
\medskip\centerline{ with $E_6$ Global Symmetry}}}
{\baselineskip=12pt
\centerline{Joseph A. Minahan\foot{minahan@physics.usc.edu} and 
Dennis Nemeschansky\foot{dennisn@physics.usc.edu}}
\bigskip
\centerline{\sl  Department of Physics and Astronomy}
\centerline{\sl University of Southern California}
\centerline{\sl Los Angeles, CA 90089-0484}
\medskip
\bigskip
\centerline{\bf Abstract}

}

\noindent
We obtain the elliptic curve corresponding to an $N=2$ superconformal field
theory which has an $E_6$ global symmetry at the strong coupling point
$\tau=e^{\pi i/3}$.  We also find the Seiberg-Witten differential 
$\lambda_{SW}$ for this theory.  This differential has 27 poles
corresponding to the fundamental representation of $E_6$.  The complex
conjugate representation has its poles on the other sheet.  We also
show that the $E_6$ curve reduces to the $D_4$ curve of Seiberg and
Witten.   Finally, we compute the monodromies and use these to compute
BPS masses in an $F$-Theory compactification.

\Date{8/96}
\vfil
\eject

\def\NP{{\it Nucl. Phys.\ }}
\def\PL{{\it Phys. Lett.\ }}

\def\PRL{{\it Phys. Rev. Lett.\ }}

\def\al{\alpha}

\def\la{\lambda}
\def\th{\theta}

\def\Imt{{\rm Im}\tau}

\def\lSW{\lambda_{SW}}

\lref\SWI{N. Seiberg and E. Witten, {\it Electric-Magnetic Duality, Monopole 
Condensation, and Confinement in $N=2$ Supersymmetric Yang-Mills Theory},
{\bf hep-th/9407087}, \NP{\bf B426} (1994) {19}.}
\lref\SWII{N. Seiberg and E. Witten, {\it Monopoles, Duality and Chiral
Symmetry Breaking in $N=2$ Supersymmetric QCD}, {\bf hep-th/9408099},
\NP {\bf B431} (1994) {484}.}
\lref\LW{W. Lerche and N. Warner, \NP{\bf B358} (1991) 571.}
\lref\Vafa{C. Vafa, {\it Evidence for $F$-Theory,} {\bf hep-th/9602022}, 
{\it Nucl. Phys.} {\bf B469} (1996) 403.}
\lref\sen{A. Sen, {\it $F$-Theory and Orientifolds}, {\bf hep-th/9605150}.}
\lref\KLTY{A. Klemm, W. Lerche, S. Theisen and S. Yankielowicz,
	{\bf hep-th/9411048}, \PL {\bf B344} (1995) {169}.}
\lref\AF{P.C. Argyres and A.E. Faraggi, {\bf hep-th/9411057},
	\PRL {\bf 73} (1995) {3931}.}
\lref\HO{A. Hanany and Y. Oz, {\it On the
	Quantum Moduli Space of Vacua of N=2 Supersymmetric $SU(N_c)$
	Gauge Theories, {\bf hep-th/9505075}, {\it Nucl. Phys.} {\bf B452} 
(1995) 283-312.}}
\lref\APS{P.~C.~Argyres, M.~R.~Plesser and A.~D.~Shapere, 
{\it The Coulomb Phase of N=2 Supersymmetric QCD}, {\bf hep-th/9505100},
{\it Phys. Rev. Lett.} {\bf 75} (1995) 1699-1702.} 
\lref\FK{H.~M.~Farkas and I.~Kra, {\it Riemann Surfaces}, Springer-Verlag 
(1980), New York.}
\lref\Clemens{C.~H.~Clemens, {\it A Scrapbook of Complex Curve Theory}, Plenum
Press (1980), New York.}
\lref\MN{J. Minahan and D. Nemeschansky, 
{\it Hyperelliptic curves for Supersymmetric Yang-Mills}, 
{\bf hep-th/9507032}, \NP{\bf 464} (1996) {3}.}
\lref\MNII{J. Minahan and D. Nemeschansky, 
{\it $N=2$ Super Yang-Mills and Subgroups of $SL(2,Z)$}, 
{\bf hep-th/9601059}, {\it to appear in Nucl. Phys. B}.}
\lref\AD{P. Argyres and M. Douglas, {\it New Phenomena in $SU(3)$ 
Supersymmetric Gauge Theory}, {\bf hep-th/9505062}, 
{\it Nucl. Phys.} {\bf B448} (1995) 93-126}
\lref\AY{O. Aharony and S. Yankielowicz, {\it Exact Electric-Magnetic Duality 
in $N=2$ Supersymmetric QCD Theories}, {\bf hep-th/9601011}.}
\lref\Kob{N. Koblitz, {\it Introduction to Elliptic Curves and Modular Forms},
Springer-Verlag, (1984), New York.}
\lref\Sch{B. Schoeneberg, {\it Elliptic Modular Functions; An Introduction},
Springer-Verlag, (1974), New York.}
\lref\FP{D. Finnell and P. Pouliot, {\it Instanton Calculations Versus Exact 
Results In 4 Dimensional Susy Gauge Theories}, {\bf hep-th/9503115}.}  
\lref\DW{R. Donagi and E. Witten, {\it Supersymmetric Yang-Mills Systems And 
Integrable Systems}, {\bf  hep-th/9510101}.}
\lref\MWI{E. Martinec and N. Warner, {\it Integrable systems and 
supersymmetric gauge theory}, {\bf hep-th/9509161}.}
\lref\Mart{E. Martinec, {\it Integrable Structures in Supersymmetric Gauge 
and String Theory}, {\bf hep-th/9510204}.}
\lref\MWII{E. Martinec and N. Warner, {\it Integrability in N=2 Gauge Theory: 
A Proof}, {\bf hep-th/9511052}.}
\lref\GKMMM{A.Gorsky, I.Krichever, A.Marshakov, A.Mironov and A.Morozov,
{\it Integrability and Seiberg-Witten Exact Solution}, {\bf hep-th/9505035},
{\it Phys. Lett.} {\bf B355} (1995) 466-47.}
\lref\NT{T. Nakatsu and K. Takasaki, {\it Whitham-Toda hierarchy and $N = 2$
 supersymmetric Yang-Mills theory}, {\bf hep-th/9509162}.}
\lref\IM{H. Itoyama and A. Morozov, {\it Integrability and Seiberg-Witten 
Theory: Curves and Period}, {\bf hep-th/9511126}.}
\lref\APSW{P.C. Argyres, M.R. Plesser, N. Seiberg and E. Witten,
{\it New $N=2$ Superconformal Field Theories in Four Dimensions}, 
{\bf  hep-th/9511154}, {\it Nucl.Phys.} {\bf B461} (1996) 71.} 
\lref\EHIY{ T. Eguchi, K. Hori, K. Ito and S.-K. Yang,
{\it Study of $N=2$ Superconformal Field Theories in $4$ Dimensions},
{\bf hep-th/9603002}, {\it Nucl. Phys.} {\bf B471} (1996) 430.}
\lref\DM{K. Dasgupta and S. Mukhi, {\it $F$-Theory at Constant Coupling},
{\bf hep-th/9606044.}}
\lref\Kod{K. Kodaira, {\it On Compact Analytic Surfaces, II-III}, {\it Ann.
of Math.} {\bf 77} (1963) 563; {\bf 78} (1963) 1.}

%          \newsec{Introduction}

\newsec{Introduction}

Kodaira's classification of the singularities of the torus demonstrates
an ADE pattern\Kod.  Singularities that occur at $\Imt=\infty$ are either
$A_n$ or $D_n$.  There are also a finite number of
singularities that occur at finite
values of $\tau$.  The singularity types for these are $A_0$, $A_1$, $A_2$,
$E_6$, $E_7$ and $E_8$.  Finally, there is a $D_4$ singularity that
can occur at all values of $\tau$.  

What is quite striking about this is that most of these singularities
have appeared in $N=2$ $U(1)$ superconformal field theories.  In
particular, $D_4$ first
appeared in the classic papers of Seiberg and Witten\refs{\SWI,\SWII} 
and $A_0$, $A_1$ and $A_2$ appear
in  certain limits of an $SU(2)$ gauge theory with $N_f=1,2$ or $3$ 
respectively\refs{\APSW,\EHIY}, 
and hence can be derived from the $D_4$ theory. 
(The $A_0$  theory also can come from an $SU(3)$ gauge
theory with $N_f=0$\AD).  
The theories at $\Imt=\infty$ are basically trivial since
the coupling runs to zero.

What are missing are the theories  with $E_6$, $E_7$ and $E_8$ singularities.
In this paper, we will take a step in this direction by constructing
the explicit elliptic curve for $E_6$, along with the corresponding
Seiberg-Witten differential.  It is not clear to us if such a theory
can be reached in some limit of a super Yang-Mills theory.  The $E_7$ and
$E_8$ cases will be discussed in a separate publication

\newsec{Dimensions and Relevant Operators}

Let us begin by showing why a superconformal fixed point is related to
Kodaira's classification.
We assume that the Seiberg-Witten curve is of the form
\eqn\torus{
y^2=x^3-f(\rho)x-g(\rho)
}
where $f(\rho)$ and $g(\rho)$ are polynomials in $\rho$ and $\rho$ is the
expectation value of some scalar field.  We further assume  
that there exists a differential $\lambda_{SW}$ such that
\eqn\SWdiff{
{d\lambda_{SW}\over d\rho}={dx\over y}.
}
Hence $\lambda_{SW}$ has the same dimensions as $\rho dx/y$.  Since BPS
masses are found by integrating $\lSW$ around closed loops, $\lSW$ has
dimension $1$.  In order to have a unitary quantum field theory, the dimension
of any operator should be nonnegative.  The type of singularity is determined
by the behavior of $f$, $g$ and the discriminant $\Delta=4f^3-27g^2$.
With no loss of generality, we can assume that the singularity occurs
at $\rho=0$ and that $f\sim \rho^r$,  $g\sim\rho^s$.  

If $2s<3r$, then the singularity is dominated by $g$.  In this case,
the dimension of $\rho$, $[\rho]$, satisfies $s[\rho]=3[x]=2[y]$. 
Since $[\lSW]=1$, one finds $[\rho]=6/(6-s)$.  Therefore, one must have
$s<6$ in order to have
a unitary theory.  At the singularity, the coupling is $\tau=e^{\pi i/3}$. 
For $s=1$ and $s=2$, the singularity is $A_0$ and $A_2$ respectively.  For
$s=3$, the singularity is $D_4$, and for $s=4,5$, the singularity is 
$E_6$ and $E_8$ respectively.  

If $2s>3r$, then the singularity is determined by
the behavior of $f$. In this case, $r[\rho]=2[x]$, hence $[\rho]=4/(4-r)$
and $m<4$ in order for the theory to be unitary. If $r=1$, the singularity 
is $A_1$, $r=2$ is $D_4$ and $r=3$ is $E_7$. 

If $2s=3r$, then
a unitary theory will either have $s=0$ or $s=3$.  The singularity then
depends on the discriminant.  In the case where $s=0$, and 
$\Delta\sim\rho^{n+1}$,
the singularity is $A_n$ and occurs at weak coupling.  If $s=3$ and 
$\Delta\sim\rho^{6+n}$, then the singularity is $D_{4+n}$ and occurs
at weak coupling for $n>0$.

Each of these theories will flow away from criticality as relevant operators
are turned on.  For what follows, we will consider only the superconformal
theories that occur at strong coupling.  The number of relevant operators
depends on the dimension of the polynomials $f$ and $g$ that will lower the
critical exponent in $\Delta$.  So for instance, the $A_2$ singularity is
reduced by adding to $g$ the polynomial $a\rho+b$ and to $f$ the polynomial
$c\rho+d$.  This has four total degrees of freedom, which leads to
three relevant operators, since one degree of freedom can be used to shift
$\rho$.  The strong coupling $A_n$ singularities will have $n+1$ relevant 
operators.

The $D_4$ theory has four relevant operators as well as a marginal operator,
namely the bare coupling.  In fact, the $D_4$ theory as well as the $A_1$ 
and $A_2$
theories each has an operator that allows one to flow from strong to
weak coupling while still preserving the global symmetry.  In the case
of $A_1$ and $A_2$, this operator can be thought of as the center of mass 
operator in the $SU(2)$ gauge theory, or equivalently as the cutoff.  

However, for the $E_n$ cases, the number of relevant operators is $n$,
as one can easily verify.  Moreover, there is no marginal operator
analogous to the bare coupling in the $D_4$ case.  Hence, in order
to flow away from the strong coupling point, one must break the global
symmetry.  Therefore, all relevant operators should be expressible in
terms of the $E_n$ casimirs.

In the rest of this paper, we will use symmetry arguments as well as
some of the ideas presented in section 17 of \SWII\ to construct
the elliptic curve for the $E_6$ case.

\newsec{Construction of the curve for $E_6$}

It is convenient to express the curve in terms of the casimirs
of the $U(1)\times SO(10)$ subgroup of $E_6$.  We define operators
corresponding to the Cartan Subalgebra of this subgroup, $\lambda$ and
$m_i$.  We also assume that the residues of $\lSW$ will be linear
combinations of these operators, hence they must all have dimension 1.
We then define the following
$SO(10)$ casimirs
\eqn\SOtencas{\eqalign{
&T_2=\sum_{i=1}^5m_i^2\qquad\qquad T_4=\sum_{i<j}m_i^2m_j^2\qquad\qquad T_6
=\sum_{i<j<k}m_i^2m_j^2m_k^2\cr
&T_8=\sum_{i<j<k<l}m_i^2m_j^2m_k^2m_l^2\qquad\qquad T_5=m_1m_2m_3m_4m_5.}
}

We now proceed incrementally.  First assume that $m_i=0$ for all $i$.  A
nonzero $\lambda$ breaks the global symmetry from $E_6$ to $SO(10)$.
Since $[\rho]=3$ and $[x]=4$, the curve must be of the form
\eqn\esixsoten{
y^2=x^3-12\lambda^2\rho^2x-\rho^4+16\rho^3\lambda^3,
}
up to an overall normalization of the $U(1)$ casimir and an overall shift
in $\rho$.  The discriminant behaves as $\rho^7$ corresponding to a
$D_5$ global symmetry.

We next turn on $m_1$, breaking the global symmetry to $SO(8)$.  The only
nonzero $SO(10)$ casimirs are made up of powers of $T_2$.  Hence the
generic curve is then
\eqn\esixsoeight{
y^2=x^3-(12\lambda^2+ T_2)\rho^2x-(\rho^4-2\lambda(8\lambda^2+\alpha T_2)).
}
The first coefficient of $T_2$ is determined by choosing an overall scale.
In order to determine the coefficient $\alpha$  we have
to assume that $\lSW$ has poles whose residues are linear combinations
of $\lambda$ and $m_i$.  We may also assume that the differential has
a factor of $y$ in the denominator.  Hence, in order to satisfy 
these requirements, $y^2$ in \esixsoeight\ must be a perfect square
when $x$ is at the position of the pole.  Following \SWII, we
assume that the poles have a linear dependence on $\rho$ and can be
written in the form
\eqn\xpol{
x=\beta\rho+\theta.
}
Clearly, $\theta$ must be a perfect square, and given the dimensions
of $\rho$, $m_1$ and $\lambda$, $x$ has the form
\eqn\xpole{
x=(am_1+b\lambda)\rho-(rm_1^2+sm_1\lambda+t\lambda^2)^2
}
At the pole, $y$ should have the form
\eqn\ypole{
y^2=-(\rho^2+B\rho+C)^2
}
Hence, from the linear and constant pieces in $y^2$, we find
\eqn\BCpole{
C=(rm_1^2+sm_1\lambda+t\lambda^2)^3\qquad\qquad 
B=-{3\over2}(am_1+b\lambda)(rm_1^2+sm_1\lambda+t\lambda^2)
}
If $\alpha=-2$, then we find four sets of solutions, which
are 
\eqn\solone{
(a,b,r,s,t)=(\pm1,2,0,0,0), (0,8,9/2,0,0), (\pm2,-4,0,\mp6, 4),(0,-4,0,0,0)
}
Recall that the fundamental representation of $E_6$ decomposes under
$SO(10)\times U(1)$ to
\eqn\fundrep{
{\bf 27}={\bf 16}_{+1}+{\bf 10}_{-2} +{\bf 1}_{+4}.}
Hence, it is clear that the first solution in \solone\ is a spinor,
the second is a singlet, and the last two are vector solutions, with
the first of these having the vector component aligned along $m_1$ and
the last solution having it orthogonal to $m_1$.  Therefore, we
expect to find poles at $x_\al=2h_\alpha\rho+\th_\al$, where $\al$ is
an index transforming in the fundamental of $E_6$.  In terms of $SO(10)$
representations, these poles are at
\eqn\xspsvpole{\eqalign{
x_{\pm\pm\pm\pm\pm}&=(\pm m_1\pm m_2\pm m_3\pm m_4\pm m_5+2\lambda)\rho
+\theta_{\pm..\pm}\cr
x_{\pm i}&= (\pm 2m_i-4\lambda)\rho+\theta_i\cr
x_s  &=(+8\lambda)\rho+\theta
}}
where $\th$, $\th_i$ and $\th_{\pm..\pm}$ are yet to be determined and the
number of $+$ signs in $x_{sp}$ is even.

In fact, we can now find the complete curve by symmetry arguments and
the assumption that  $y^2$ is a perfect square when $x$ is a spinor
solution in \xspsvpole.  In particular, if $m_3=m_4=m_5=0$, there is
a remaining $SO(6)\simeq SU(4)$ global symmetry, hence the discriminant
should behave as $\rho^4$ as $\rho\to 0$.  If $m_4=m_5=0$, then there
is an $SO(4)\simeq SU(2)\times SU(2)$ global symmetry, in which case
the discriminant behaves as $(\rho^2-\gamma^2)^2$, where $\gamma$ is
some constant.

The final result for the curve is
\eqn\curve{\eqalign{
y^2=x^3&-\biggl(\rho^2(12\lambda^2+T_2)+\rho(8T_5+8\lambda T_4)+{1\over3}T_4^2+
(36\la^2-T_2)T_6+4T_8-12\la(36\la^2-T_2)T_5\biggr)x\cr
&-\biggl(\rho^4-4\rho^3\la(4\la^2-T_2)+\rho^2\Bigl({1\over3}T_2T_4-2T_6
+20\la^2T_4-40\la T_5\Bigr)\cr
&\qquad+\rho\Bigl({8\over3}\la T_4^2-4\la T_6(T_2-36\la^2)
-{16\over3}T_4T_5-32\la T_8+2T_2^2T_5-96\la^2 T_2T_5+864\la^4T_5\Bigr)\cr
&\qquad
+\Bigl({2\over27}T_4^3-{1\over3}T_2T_4T_6+T_6^2+(T_2-36\la^2)^2T_8-
{8\over3}T_4T_8+12\la^2T_4T_6+144\la^2T_5^2\cr
&\qquad\qquad\qquad-24\la T_5T_6-144\la^3 T_4T_5+4\la T_2T_4T_5\Bigr)\biggr)
}
}
This curve is a perfect square at the spinor point
\eqn\xsp{
x_{+++++}=(T_1+2\la)\rho-{1\over3}T_4+2T_4'-(6\la+T_1)T_3,}
where
\eqn\oddTs{
T_1=\sum_i m_i,\qquad\qquad T_3=\sum_{i<j<k}m_im_jm_k\qquad\qquad 
T'_4=\sum_{i<j<k<l}m_im_jm_km_l.
}
\curve\ is still a perfect square if we change the signs of an even
number of the $m_i$ in $T_1$, $T_3$ and $T_4$.
For  $x=x_{+++++}$ in \xsp, $y^2$ satisifies
\eqn\ysp{\eqalign{
y^2=y^2_{+++++}&=-\biggl(\rho^2+{1\over2}(T_1+6\la)(T_1^2-T_2)(T_3-\rho)-T_6\cr
&\qquad\qquad-T_4'(2T_1^2-T_2-18\la(T_1+2\la))+4T_5(T_1+12\la)\biggr)^2.}
}
The vector points are given by
\eqn\xv{\eqalign{
x_{\pm i}&=2(\pm m_i-2\la)\rho+m_i^2\Bigl(-m_i^2-(\pm m_i-6\la)^2+T_2\Bigr)
-{1\over3}T_4\pm 2T_5/m_i\cr
y^2_{\pm i}&=-\biggl(\rho^2-\Bigl(4m_i^2(\pm m_i-6\la)\pm m_i(36\la^2-T_2)
\Bigr)\rho+3m_i^6\mp6\la m_i^5-3m_i^4T_2\pm6\la m_i^3T_2\cr
&\qquad\qquad
+2m_i^2T_4-T_6\mp(4m_i^2\mp12\la m_i+36\la^2-T_2)/m_i\pm m_i^3(\pm m_i-6\la)^3\biggr)^2}
}
and the singlet point is
\eqn\xs{\eqalign{
x_{s}&=8\la\rho-{1\over4}(T_2-36\la^2)^2+{2\over3}T_4\cr
y^2_s&=-\left(\rho^2+6\la(T_2-36\la^2)\rho
+{1\over8}(4T_4-(T_2-36\la^2)^2)(T_2-36\la^2)-T_6+12\la T_5\right)^2,}
}

Eq. \curve\ should also be expressible in terms of $E_6$
casimirs.  At first, this might seem problematic, since the curve has
dimension three operators, but no such $E_6$ casimir exists.  However,
since $\rho$ is also dimension three, we can remove this term by 
shifting $\rho$ by $\la(36\la^2-T_2)$.  In terms of $E_6$ casimirs,
the new curve is
\eqn\curveesix{\eqalign{
y^2=x^3&-\biggl(-{1\over3}\rho^2P_2+{2\over3}\rho P_5-{7\over432}P_2^4
+{11\over45}P_2P_6-{8\over15}P_8\biggr)x\cr
&-\biggl(\rho^4+\rho^2\Bigl({2\over3}P_6-{7\over108}P_2^3\Bigr)
+\rho\Bigl({1\over18}P_2^2P_5-{8\over21}P_9\Bigr)\cr
&\qquad+{32\over135}P_{12}-{298\over18225}P_2^2P_8-{101\over218700}P_2^3P_6
+{13\over405}P_6^2-{49\over1049760}P_2^6-{19\over3645}P_2P_5^2\biggr),
}
}
where the $P_i$ are the $E_6$ casimirs found in \LW\foot{
Note that the relation between $\la$ and the  
term $x_1$ which appears in \LW\ is $x_1=\la$.}

\newsec{The Seiberg-Witten Differential for $E_6$}

Once the positions of the poles and their residues are known, one
can find the full differential $\lSW$.  The differential is just
a sum over the 27 of $E_6$ plus a piece that has no pole and is
hence proportional to the holomorphic differential $dx/y$.  This last
piece should be invariant under $E_6$.

When a closed curve crosses a pole under a monodromy transformation, 
the coordinate corresponding to the integral of $\lSW$ along that curve
shifts by the residue of that pole multiplied by $2\pi i$.
Therefore, $\lSW$ should be of the form
\eqn\swdiff{
\lSW=\gamma \left(\rho+\la T_2-4\la^3\right){dx\over y}+
{1\over2\sqrt{2}\pi i}\sum_{\alpha=1}^{27}
{h_\alpha y_\alpha\over x-2h_\alpha\rho-\th_\alpha}{dx\over y},
}
where $h_\alpha$ is the linear combination of $m_i$ and $\la$ for
the $\alpha$ state in the representation, $y_\alpha$ is the value of
$y$ when $x$ is at the pole and $\gamma$ is a constant to be determined.
Notice that the residues in \swdiff\ switch sign when moving to the other
sheet.  These poles then transform in the ${\bf\overline{27}}$ representation.

The constant $\gamma$ can be determined by finding $d\lSW/d\rho$ and making sure 
that it is proportional to $dx/y$, up to a total derivative.  Setting
$m_2=m_3=m_4=m_5=0$, simplifies the calculation, and one finds that
\eqn\dsw{
\gamma={9\over\sqrt{2}\pi}, \qquad\qquad 
{d\lSW\over d\rho}={3\over\sqrt{2}\pi}{dx\over y} + {d(...)\over dx}
}

\newsec{Relation to the $D_4$ case of Seiberg and Witten}

If we take $\la$ and one of the $m_i$ to infinity, leaving the other
$m_i$ finite, the curve in \curve\ should reduce to the Seiberg-Witten
result\SWII.  This then will provide a useful check on our result.

In fact, the proper scaling is quite simple.  Choose $\la=-c_1\Lambda/6$,
$m_5=-c_2\Lambda$, $\rho=u\Lambda$ and scale $x\to x\Lambda^2$ and
$y\to y\Lambda^3$, where $c_1$ and $c_2$ are defined in \SWII.  
Keeping only the leading terms in $\Lambda$ and  shifting $x$ 
by $(c_1u+c_2^2t_2)/3$, where
$t_2=m_1^2+m_2^2+m_3^2+m_4^2$, the curve in \curve\ reduces to
\eqn\swcurve{\eqalign{
y^2&=(x^2-c_2^2u^2)(x-c_1u)-c_2^2(x-c_1u)^2t_2-c_2^2(c_1^2-c_2^2)(x-c_1u)
t_4\cr
&\qquad\qquad+2c_2(c_1^2-c_2^2)(c_1x-c_2^2u)t_4'-c_2^2(c_1^2-c_2^2)^2t_6,}
}
where
\eqn\tdefs{
t_4=\sum_{i<j<5}m_i^2m_j^2\qquad\qquad t_4'=m_1m_2m_3m_4\qquad\qquad
t_6=\sum_{i<j<k<5}m_i^2m_j^2m_k^2.
}
This is the Seiberg-Witten result\SWII.

We can also study the behavior of $\lSW$ in this limit.  To leading order
in $\Lambda$, the poles behave as
\eqn\scaledpoles{\eqalign{
x_s&=-{1\over4}(c_1^2-c_2^2)\Lambda^2\cr
x_{\pm5}&=c_2^2(c_1\pm c_2)^2\Lambda^2\cr 
x_{\pm i}&=c_1u+m_i^2(c_2^2-c_1^2)\cr
x_{\pm\pm\pm\pm+}&=c_2u+(c_1+c_2)c_2\sum_{i<j<5}(\pm m_i)(\pm m_j)\cr
x_{\pm\pm\pm\pm-}&=-c_2u-(c_1-c_2)c_2\sum_{i<j<5}(\pm m_i)(\pm m_j).}
}
Hence, the poles at $x=x_s, x_{\pm5}$ move out to infinity where the sum
of their residues cancel.  The other poles are at finite values of $x$.
However, they all have infinite residues.  But these infinite parts will
cancel off since there are also poles coming from the ${\bf\overline{27}}$. 
In the limit of $\Lambda\to\infty$, each pole in the $\bf{27}$ moves
to the same point on the torus as a corresponding pole in the $\bf{\overline{27}}$.
The infinite pieces cancel off, leaving a residue that is twice the residue
in \SWII.
Moreover, the poles split into the vector, spinor and spinor bar representation
of $SO(8)$.  By triality, the poles of any one representation are enough
to describe $\lSW$.  Hence, $\lSW$ for $E_6$ flows to $\lSW$ for $SO(8)$,
but multiplied by a factor of 6.  

\newsec{Monodromies and Applications to $F$-Theory}

The curve in \curve\ has eight singularities in the $\rho$ plane.  Since
$E_6$ has an $SU(6)\times SU(2)$ subgroup, we expect to find an $A_5$
singularity at weak coupling for some values of the $m_i$ and $\la$.  
Hence the monodromies around six of the singularities, $N_i$, commute with
each other.  Indeed, the $A_5$ singularity occurs if $m_i=m_j=-2\la=m$.
In this case, the discriminant is proportional to
\eqn\sixdis{
\Delta\sim (\rho-2m^3)^6(27\rho^2+256m^6).
}

As in \SWII, the monodromies around the other two singularities $M_1$ 
and $M_2$, do not
commute with the monodromies around these six singularities, nor with
each other.  Because $\lSW$ has poles, the monodromies are not in
$SL(2,Z)$.  However, let us first ignore the contributions of the
poles.  From \sixdis, we expect $N_i=T$
and $M_1$ and $M_2$ to be conjugate to
$T=\left(\matrix{1&1\cr0&1}\right)$.  Since $\rho$ has the topology of
the sphere, we should also have that
\eqn\moneq{
M_2M_1 \prod_i N_i=M_\infty^{-1}.}
The monodromy at $\infty$ is not the same as the $D_4$ case, but is instead
\eqn\Minf{
M_\infty=T^{-1}ST^{-1}S,\qquad\qquad S=\left(\matrix{0&-1\cr1&0}\right).
}
It is not hard to show that when $m_i=\la=0$, \moneq\ is satisfied if
\eqn\Mmon{\eqalign{
M_1&=(STS)^{-1}T(STS)=-ST^{-2}\cr
M_2&=(STST^{1}S)^{-1}T(STST^{-1}S)=M_\infty M_1 M_\infty^{-1}.}
}

If we now turn on $m_i$ and $\la$, then the monodromies are modified,
since in going around a singularity, a closed loop can cross a pole.
We have some freedom to choose how the monodromies
are modified.  A convenient basis leaves $M_\infty$ unchanged and modifies
the other monodromies such that
\eqn\Nmon{\eqalign{
N_i\left(\matrix{a_D\cr a}\right)&=\left(\matrix{a_D+a+m_i-2\la\cr a}\right)
\qquad\qquad 1\le i\le5\cr
N_6\left(\matrix{a_D\cr a}\right)&=\left(\matrix{a_D+a+{1\over2}\sum_i m_i +\la
\cr a}
\right)}
}
\eqn\Mmonm{\eqalign{
M_1\left(\matrix{a_D\cr a}\right)&=\left(\matrix{a+
{1\over2}(m_1+m_2+m_3-m_4-m_5)-3\la\cr
-a_D+2a+m_5-6\la}\right)\cr
M_2\left(\matrix{a_D\cr a}\right)&=\left(\matrix{3a_D+a+3m_4+2m_5+6\la\cr
-4a_D-a+{1\over2}(m_1+m_2+m_3-7m_4-5m_5)-\la}\right)}
}
One can easily check that \Minf, \Nmon\ and \Mmonm\ satisfy \moneq.

These results can be used to investigate some aspects of $F$-Theory\Vafa.
In particular, Sen has argued that one can use the Seiberg-Witten
result to study $F$-Theory compactified on the $T_4/Z_2$ orbifold\sen.
He argued that this is equivalent to a type IIB theory on an orientifold.
At the orbifold point, there is an enhanced $(SO(8))^4$ gauge theory.
Sen then uses the results of \SWII\ to see what happens as one moves 
away from the orbifold point.  At the orbifold point, the 24 singularities
in the $\rho$ plane break up into 4 groups of 6.  One moves away from
the orbifold point by resolving the singularities, which breaks the
gauge symmetry.  

In particular, if
we pick one group of six singularities, then the BPS masses are given
by integrals of $\partial_\rho a_D$ along closed loops in the $\rho$ plane.
So one finds a BPS mass $m_i-m_j$ by integrating counterclockwise around
the $i$th singularity and clockwise around the $j$th singularity.  In the
type IIB language, this is equivalent to stretching an open string between
the $i$th and $j$th $D$-brane.  One can get a BPS mass $m_i+m_j$ by integrating  
counterclockwise around the $i$th singularity, then counterclockwise around
the outside of the fifth and sixth singularity, clockwise around the
$j$th singularity and back clockwise around the fifth and sixth singularity.
For IIB, this is equivalent to having an open string stretched between
the $i$th $D$-brane and an orientifold plane and back again to the $j$th
$D$-brane.  In IIB perturbation theory, there are two orientifold planes
at the same location, however, from Sen's analysis, we see that quantum
effects should split these apart.  However, this has no consequences
for the BPS states since we only integrated around both singularities
and never each one individually.

In \DM, the authors considered $F$-theory compactified on other orbifolds.
However, unlike the $T_4/Z_2$ case, these orbifolds only exist
for particular values of the toroidal moduli.  The orbifold is
an elliptical fibration over the $\rho$ plane and the coupling is
given by the modulus of the elliptic curve, hence these theories must
be strongly coupled.  The $T_4/Z_3$ orbifold should
have an enhanced $E_6\times E_6\times E_6$ gauge theory, where the
$\rho$ plane has 3 groups of 8 singularities.  

It is not clear how to  
think of this theory in terms of type IIB.  But we can use the monodromies
in \Minf, \Nmon\  and \Mmon to find integrals of $\partial_\rho a_D$ along
closed paths in the $\rho$ plane that give the correct masses for the BPS
states for the broken $E_6$ gauge theory.
The adjoint rep of $E_6$ decomposes into a $SO(10)\times U(1)$ subgroup as
\eqn\adj{
{\bf 78}={\bf 45}_0+{\bf 1}_0+{\bf 16}_{-3}+{\bf\overline{16}}_{+3}.
}
It is clear from the monodromies in \Nmon\ that integrating counterclockwise
around the $i$th singularity and clockwise around the $j$th singularity $i,j<6$
gives a mass $m_i-m_j$.  Likewise, integrating counterclockwise around the 
sixth singularity and clockwise around the $i$th singularity gives a mass
that comes from the ${\bf\overline{16}}$ representation.  But not all
states in the ${\bf\overline{16}}$  are found this way.

To find the other masses, note that if $m_i=\la=0$, then $M_1N_iN_j=
M_\infty N_k^{-1}=S^{-1}$.  Hence the product of any four of these
loops is the identity.  Once the masses are turned back on, the product
of any four of these shifts $a_D$ by a piece linear in the $m_i$ and
$\la$.  For instance,
\eqn\loopi{\eqalign{
M_1N_pN_q&M_\infty N_k^{-1}M_\infty N_j^{-1} M_\infty N_i^{-1}\left(\matrix{
a_D\cr a}\right)\cr
&=\left(\matrix{a_D+m_k-m_i+{1\over2}(m_1+m_2+m_3-m_4-m_5)-3\la
\cr
a-m_j-m_p-m_q+m_5}\right)\qquad\qquad1\le i,j,k,p,q\le5}
}
so this then leads to BPS masses for other states in the ${\bf 16}$ and
${\bf\overline{16}}$.  Using \moneq, we can replace $M_\infty$ with
$N_i$ and $M_j$.  To get the missing states in the ${\bf 45}$, we use
the fact that
\eqn\loopii{\eqalign{
M_\infty N_k^{-1}&M_1N_pN_qM_\infty N_j^{-1} M_\infty N_i^{-1}\left(\matrix{
a_D\cr a}\right)\cr
&=\left(\matrix{a_D-m_i-m_p-m_q+m_5\cr
a+m_k-m_j+{1\over2}(m_1+m_2+m_3-m_4-m_5)-3\la}\right)\qquad\qquad
1\le i,j,k,p,q\le5.}
}
Hence, setting $i=5$ and integrating $\partial_\rho a_D$ over this path
gives $-m_i-m_j$.  Finally, the state with mass $\sum m_i-3\la$ can
be obtained by combining the loops in \loopi\ and \loopii.  

Obviously, these loops are more complicated than the loops in \sen.  Moreover,
in order to get the complete set of BPS states it is necessary to loop around
the seventh and eighth singularities individually.  This suggests
that strong coupling monodromies are important in finding the masses for the
vector bosons.  Hopefully, this analysis will lead to a better understanding
of the type IIB string dynamics.

{\bf Acknowledgements}: 
D.N. would like to thank the Aspen Center for Physics for hospitality 
during the completion of this work.
This research was supported in part by D.O.E.~grant DE-FG03-84ER-40168.
%\newsec{Discussion}

\listrefs

\bye